\documentclass[
    ,final            
  ,sort&compress      
  ]
  {aipproc}

\layoutstyle{6x9}
\newcommand{\open}{{<\kern -0.3 em{\scriptscriptstyle )}}}

\begin{document}

\title{Transversity and inclusive two-pion production}



\author{Marco Radici}{
  address={Dipartimento di Fisica Nucleare e Teorica, Universit\`a di Pavia 
  and \\
  Istituto Nazionale di Fisica Nucleare e Teorica, Sezione di Pavia, I-27100 
  Pavia, Italy}
}

\begin{abstract}
A model for dihadron fragmentation functions is briefly outlined, that describes
the fragmentation of a quark in two unpolarized hadrons. The parameters are
tuned to the output of the PYTHIA event generator for two-hadron semi-inclusive
production in deep inelastic scattering at HERMES. Then predictions are made for
the unknown polarized fragmentation function and the related single-spin 
asymmetry in the azimuthal distribution of $\pi^+ \pi^-$ pairs in semi-inclusive
deep inelastic scattering on transversely polarized targets at HERMES and
COMPASS. This asymmetry can be used to extract the quark transversity
distribution.
\end{abstract}

\maketitle


Dihadron Fragmentation Functions (DiFF) describe the probability that a quark
hadronizes into two hadrons plus anything else, i.e.\ the process 
$q\to h_1\,h_2\,X$~\cite{tutti1}. They can appear in lepton-lepton, 
lepton-hadron and hadron-hadron collisions producing final-state hadrons. 
DiFF can be used as analyzers of the spin of the fragmenting quark. At present, 
the most important application of polarized DiFF is the measurement of the 
quark transversity distribution $h_1$ in the nucleon, which represents the 
probabilistic distribution of transversely polarized partons inside transversely 
polarized hadrons. 

Transversity is a missing cornerstone to complete the knowledge of the 
leading-order (spin) structure of the nucleon (for a review, see
Ref.~\cite{book}). Its peculiar behavior under evolution represents a basic test 
of QCD in the nonperturbative domain. The most popular strategy to extract $h_1$ 
is to consider Deep Inelastic Scattering (DIS) of electrons on transversely 
polarized targets, and look for azimuthally asymmetric distributions of 
inclusively produced single pions when flipping the spin of the target (the 
socalled Collins effect~\cite{collins}). But the cross section must
explicitly depend on the transverse momentum of the pion~\cite{boermuld}.

\begin{figure}[h]
  \centerline{
  \includegraphics[height=3.7cm]{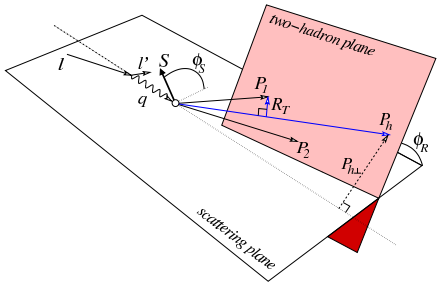} \hspace{1cm} 
  \includegraphics[width=6.5cm,height=4cm]{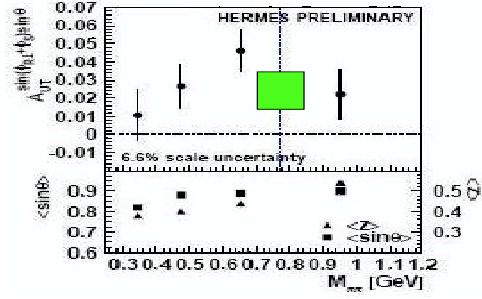}
  }
  \caption{Left panel: kinematics for deep-inelastic production of two hadrons 
  off a tranversely polarized target. Right panel: two-hadron invariant mass
  dependence of the related single-spin asymmetry, calculated in the model of
  Ref.~\protect{\cite{noiold}} adjusted to the Trento 
  conventions~\protect{\cite{trento}}, compared with the recent HERMES 
  data~\protect{\cite{hermes}}.}
  \label{fig:kin}
\end{figure}

Inclusive production of two pions offers an alternative and easier framework, 
where the chiral-odd partner of $h_1$ is represented by the DiFF $H_1^{\open}$, 
which relates the transverse spin of the quark to the azimuthal orientation of 
the $(\pi \pi)$ plane. In fact, the leading-twist spin asymmetry is~\cite{noiold}
\begin{equation}
A_{UT}^{\sin(\phi_R^{} + \phi_S^{})} \equiv  
\frac{1}{\sin (\phi_R^{} + \phi_S^{})}\, 
\frac{d\sigma^\uparrow - d\sigma^\downarrow}
     {d\sigma^\uparrow + d\sigma^\downarrow} 
\propto 
\frac{\sum_q e_q^2\,(h_1^q(x)/x)\ H_{1}^{\open q}(z,M_h^2)}
{\sum_q e_q^2\,(f_1^q(x)/x)\  D_{1}^q(z,M_h^2)} \; ,
\label{eq:asy}
\end{equation}
where $M_h$ is the invariant mass of the two pions carrying a total $z$ momentum
fraction, and $\phi_R$ and $\phi_S$ are the azimuthal orientations with respect
to the scattering plane of the $(\pi \pi)$ plane and target spin, respectively
(see Fig.~\ref{fig:kin}, left panel). 

So far, in the literature only two papers addressing $A_{UT}$ were available: one
predicting a sign change around $M_h \sim m_\rho = 770$ MeV~\cite{jaffe}, one
predicting a stable small asymmetry~\cite{noiold}. In the right panel of 
Fig.~\ref{fig:kin}, I show results for the latter one, adjusted to the
Trento conventions~\cite{trento} and compared with the recent preliminary data 
from HERMES~\cite{hermes}. The applicability range in $M_h$ was restricted around
the $\rho$ mass, and a large uncertainty band was deriving from modelling $h_1$ 
in Eq.~(\ref{eq:asy}) as well as from the unavailability of constraints on the
model parameters. Here, I present some results from a significant
upgrade of this model~\cite{noinuovo}, where in the same framework of the
"spectator" approximation a $(\pi^+ \pi^-)$ pair can be produced via a background
contribution, or via the $\rho$ resonance, or via the $\omega$ resonance decaying
either directly in $(\pi^+ \pi^-)$ or in $(\pi^+ \pi^- \pi^0)$ (and then summing
upon the unobserved $\pi^0$). The asymmetry is generated by the interference
between the first channel, where the $(\pi^+ \pi^-)$ is assumed to be produced in
a relative $s$ wave, and the other ones where the $(\pi^+ \pi^-)$ is assumed in
a relative $p$ wave. At variance with Ref.~\cite{noiold}, the parameters are 
fixed by reproducing the $M_h$ and $z$ dependences of $(\pi^+ \pi^-)$, as they 
come out from PYTHIA adjusted to HERMES kinematics. In other words, the $D_1$ in
Eq.~(\ref{eq:asy}) is fitted and the calculation of $H_1^{\open}$ is a true 
predictions.
\vspace{-0.3cm}
\begin{figure}[h]
    \centerline{
    \includegraphics[height=4.5cm]{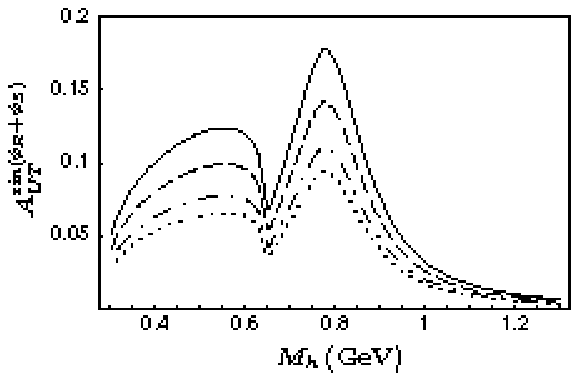} \hspace{1cm}
    \includegraphics[height=3.6cm]{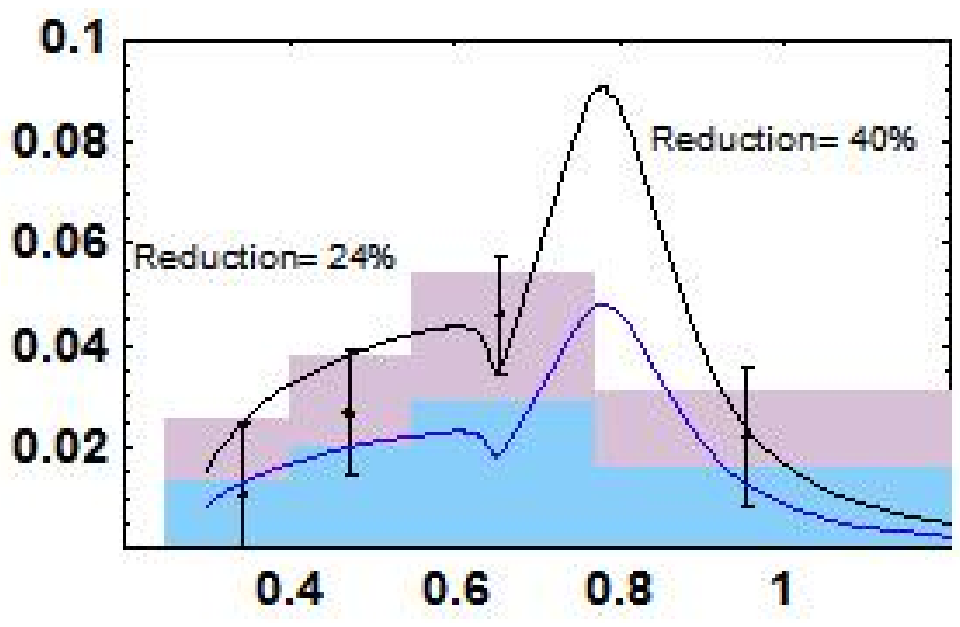}
    }
    \caption{Left panel: model results from Ref.~\protect{\cite{noinuovo}} for 
    two-pion invariant mass dependence of the single-spin asymmetry in the 
    HERMES kinematics; dotted line using transversity from
    Ref.~\protect{\cite{wak}}, dot-dashed from Ref.~\protect{\cite{nowak}},
    dashed from Ref.~\protect{\cite{peter}}, solid from
    Ref.~\protect{\cite{werner}}. Right panel: the solid lines for the results of
    Ref.~\protect{\cite{wak}} and~\protect{\cite{werner}} are transformed in 
    histograms after averaging on the experimental bins, and are fitted the 
    experimental HERMES data~\protect{\cite{hermes}}.}
    \label{fig:ssa}
\end{figure}

The results are shown in Fig.~\ref{fig:ssa}, left panel, for several models of
$h_1$ in Eq.~(\ref{eq:asy}). The statistical uncertainty on $H_1^{\open}$, coming
from the parameter fixing, is negligible~\cite{noinuovo}; the theoretical 
uncertainty comes entirely from modelling $h_1$. The shape of $A_{UT}$ is quite
good, but the overall size is too big. There are mainly two reasons for that. In
the calculations, the spectator was assumed the same in all channels, which
maximizes the interference of pion pairs in different channels~\cite{noinuovo}.
Moreover, it is known that the channel $\omega \to \pi^+ \pi^- \pi^0$ contains a
significant percentage of $(\pi^+ \pi^-)$ in relative $s$ waves; hence also this
channel has been overestimated~\cite{noinuovo}. To have an idea of the size of
the overestimation, in the right panel of Fig.~\ref{fig:ssa} the lowest and the
highest $A_{UT}$ of the left panel have been transformed in histograms, by
averaging the theoretical result over each of the five experimental bins (see 
Ref.~\cite{noinuovo} for details), and then the histograms have been fitted to 
the HERMES data. It turns out that it is necessary to reduce to 40\% the number
of $(\pi^+ \pi^-)$ pairs active in the $s-p$ interference; the number of
pairs in $p$ wave, coming from the $\omega \to \pi^+ \pi^- \pi^0$ channel, must
be further reduced to 60\%, which amounts to a total reduction of 24\% of pairs
active in this channel.

\begin{figure}[h]
    \includegraphics[height=5cm]{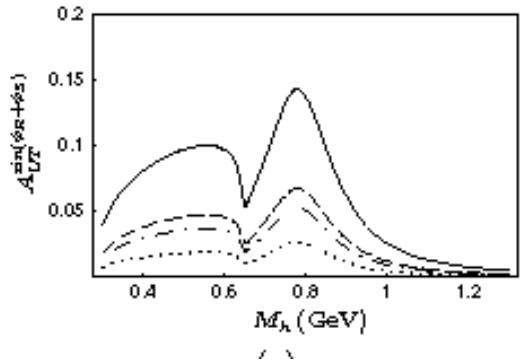}
    \caption{The same results as in Fig.~\protect{\ref{fig:ssa}}, left panel, 
    but in the COMPASS kinematics.}
    \label{fig:ssa-p}
\end{figure}

In Fig.~\ref{fig:ssa-p}, the same $A_{UT}$ is displayed in the COMPASS
kinematics, where the lepton-proton scattering c.m. energy is now $s\sim 300$
GeV$^2$ and the phase space has been reduced to $0.03<x<0.4$. 

\begin{figure}[h]
    \centerline{
    \includegraphics[height=4.5cm]{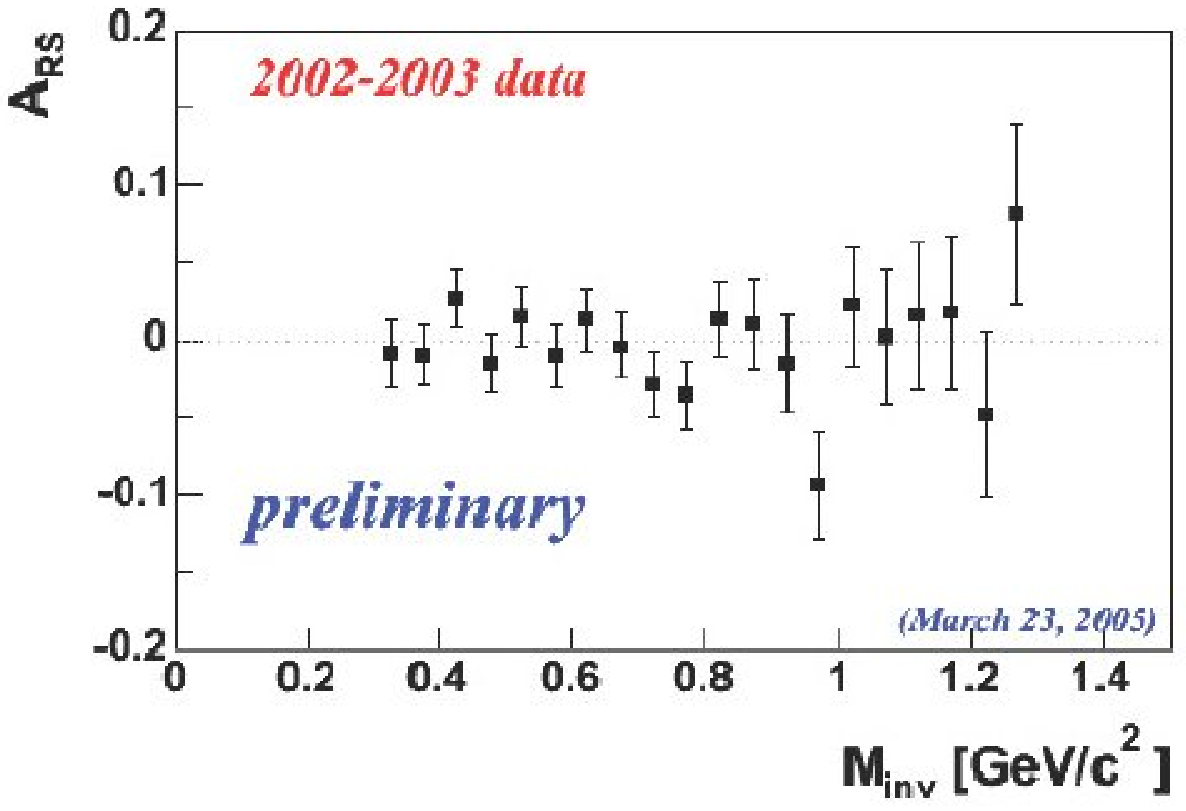} \hspace{0.1cm}
    \includegraphics[height=4.2cm]{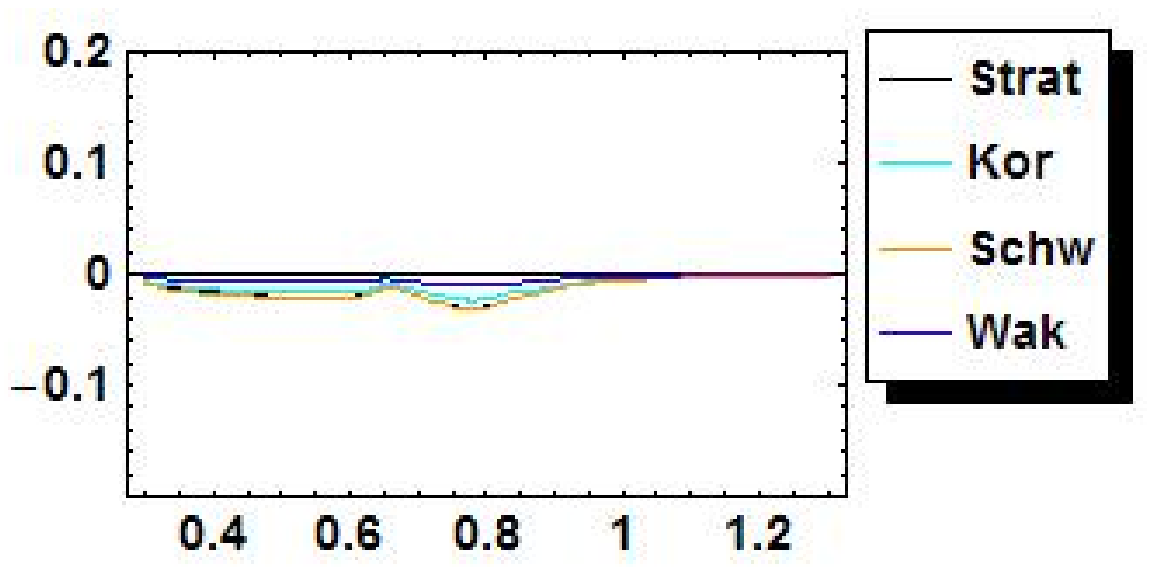}
    }
    \caption{Left panel: COMPASS data for the pion pair invariant mass dependence
    of the single-spin asymmetry for deep-inelastic scattering on a deuteron
    target~\protect{\cite{compass}}. Right panel: the same results as in 
    Fig.~\protect{\ref{fig:ssa-p}}, but for a deuteron target.}
    \label{fig:ssa-d}
\end{figure}

In Fig.~\ref{fig:ssa-d}, the COMPASS experimental data using deuteron targets
(left panel) are compared with $A_{UT}$ from Eq.~(\ref{eq:asy}) adjusted to a
deuteron target: the cancellation induced by the isospin structure of the target
is confirmed also for two-hadron inclusive production. 

Finally, it is worth to mention that there is a valid alternative to extract $h_1$
via dihadron fragmentation functions, which is of interest for all the facilities
devoted to hadronic collisions, like RHIC, GSI and, particularly, JPARC. Leaving
the details for the interested reader to Ref.~\cite{noipp}, here I just sketch the
argument. In the $pp^\uparrow \to (\pi \pi) X$ process, where a pion pair is
detected in one jet, the polarized cross section is proportional to the 
convolution
\begin{equation}
d\sigma_{UT} \propto  f_1^a \, \otimes \, h_1^b \, \otimes \, d\Delta 
\hat{\sigma}_{ab^\uparrow \to c^\uparrow d} \, \otimes \, H_1^{\open c} \; ,
\label{eq:pp-pol}
\end{equation}
where $d\Delta \hat{\sigma}$ is the elementary cross section at the partonic level.
There are two unknowns in this formula. But if in the same experiment also the 
$pp \to (\pi \pi) (\pi \pi) X$ process is considered where two pion pairs are 
detected each one in a separate jet, the cross section contains the convolution 
\begin{equation}
f_1^a \otimes f_1^b \otimes d\Delta\hat{\sigma}_{ab\to c^\uparrow d^\uparrow} 
\otimes H_1^{\open\,c} \otimes H_1^{\open \, d} \; .
\label{eq:pp-unpol}
\end{equation}
Therefore, a combined measurement of the two processes makes it possible to
self-consistently determine the unknowns, namely $H_1^{\open}$ and $h_1$.


\begin{theacknowledgments}
The collaboration with A.~Bacchetta is gratefully and warmly acknowledged.
\end{theacknowledgments}

\end{document}